\documentclass{article}

\usepackage{cite}
\usepackage{graphicx}
\usepackage{dcolumn}
\usepackage{lscape}

\begin{document}

\date{}
\title{Application of the Kronecker product to simple spin systems}
\author{Francisco M. Fern\'{a}ndez \thanks{%
E-mail: fernande@quimica.unlp.edu.ar} \\
INIFTA (CONICET, UNLP), Divisi\'on Qu\'imica Te\'orica\\
Blvd. 113 S/N, Sucursal 4, Casilla de Correo 16, 1900 La Plata, Argentina}
\maketitle

\begin{abstract}
We show that the well known Kronecker product is a suitable tool for the
construction of matrix representations of widely used spin Hamiltonians. In
this way we avoid the explicit use of basis sets for the construction of the
matrix elements. As illustrative examples we discuss two isotropic models
and an anisotropic one.
\end{abstract}

\section{Introduction}

\label{sec:intro}

In a recent paper we discussed two different operator products in quantum
mechanics\cite{F16}. We showed that one of them, the Kronecker or direct
product, is particularly useful for treating the coupling of spin systems
and for the study of the structure of NMR lines. In this paper we are
interested in the application of this product to another spin model of
practical utility.

The Breit-Rabi Hamiltonian provides a reasonable description of the
behaviour of a one-electron atom in a magnetic field. Earlier pedagogical
articles focussed on the crossings and avoided crossings between pairs of
energy levels predicted by this model\cite{DW91,B07}. The purpose of this
paper is to illustrate the utility of the Kronecker product in the
construction of the matrix representation of the Breit-Rabi Hamiltonian as
well as of other spin models.

In section~\ref{sec:Kronecker} we outline the necessary equations based on
the Kronecker product. In sections \ref{sec:isotropic1} and \ref
{sec:isotropic2} we apply the technique to two isotropic models with nuclear
spin $1/2$ and $3/2$, respectively. In section~\ref{sec:anisotropic} we
briefly consider an anisotropic model and in section~\ref{sec:conclusions}
we summarize the main results and draw conclusions.

\section{The Kronecker product}

\label{sec:Kronecker}

Here we just summarize that part of the direct product that is relevant for
present purposes. For further details the reader is referred to our earlier
paper\cite{F16}. The Kronecker product $\mathbf{A}\otimes \mathbf{B}$ of an $%
N\times N$ matrix $\mathbf{A}$ and an $M\times M$ matrix $\mathbf{B}$ is an $%
NM\times NM$ matrix with elements $A_{ni}B_{mj}$, $n,i=1,2,\ldots N$, $%
m,j=1,2,\ldots ,M$. In order to construct the Kronecker product $\mathbf{A}%
\otimes \mathbf{B}$ we just follow a simple and straightforward rule:
substitute $A_{ij}\mathbf{B}$ for every element $A_{ij}$ of $\mathbf{A}$;
for example:
\begin{equation}
\left(
\begin{array}{ll}
A_{11} & A_{12} \\
A_{21} & A_{22}
\end{array}
\right) \otimes \left(
\begin{array}{ll}
B_{11} & B_{12} \\
B_{21} & B_{22}
\end{array}
\right) =\left(
\begin{array}{llll}
A_{11}B_{11} & A_{11}B_{12} & A_{12}B_{11} & A_{12}B_{12} \\
A_{11}B_{21} & A_{11}B_{22} & A_{12}B_{21} & A_{12}B_{22} \\
A_{21}B_{11} & A_{21}B_{12} & A_{22}B_{11} & A_{22}B_{12} \\
A_{21}B_{21} & A_{21}B_{22} & A_{22}B_{21} & A_{22}B_{22}
\end{array}
\right) .  \label{eq:AxB}
\end{equation}

\section{First isotropic model}

\label{sec:isotropic1}

The Breit-Rabi formula\cite{BR33} for the interaction between an electron
and a nucleus in a magnetic field is commonly derived from the Hamiltonian
\begin{equation}
H_{BR}=A\mathbf{I}\cdot \mathbf{S}+B\left( aS_{z}+bI_{z}\right) ,
\label{eq:H_BR}
\end{equation}
where $I$ and $S$ are the nuclear and electronic spins, respectively, the
magnetic field of intensity $B$ is supposed to be along the $z$ axis, $A$ is
a measure of the coupling between the nuclear and electronic magnetic
moments and the constants $a$ and $b$ are related to the electronic and
nuclear gyromagnetic ratios and the Bohr magneton\cite{DW91,B07}.

In order to obtain a suitable matrix representation of the Hamiltonian (\ref
{eq:H_BR}) most authors resort to a basis set given by the direct product of
nuclear and electronic spin eigenvectors\cite{BR07,B07,OHPK08}
\begin{equation}
\left| m_{S},m_{I}\right\rangle =\left| S,m_{S}\right\rangle \otimes \left|
I,m_{I}\right\rangle ,\;m_{S}=-S,-S+1,\ldots ,S,\;m_{I}=-I,-I+1,\ldots ,I.
\label{eq:basis_set}
\end{equation}
In this expression $\left| S,m_{S}\right\rangle $ is an eigenvector of $%
S^{2} $ and $S_{z}$ and $\left| I,m_{I}\right\rangle $ an eigenvector of $%
I^{2}$ and $I_{z}$. In order to calculate the matrix elements $\left\langle
m_{S},m_{I}\right| \mathbf{I}\cdot \mathbf{S}\left| m_{S}^{\prime
},m_{I}^{\prime }\right\rangle $ one can, for example, express the operators
$I_{x}$, $I_{y}$, $S_{x}$ and $S_{y}$ in terms of ladder operators. However,
the straightforward application of the Kronecker product appears to be
simpler as shown in what follows.

The matrix representation of the operator $H_{BR}$ is straightforwardly
given by
\begin{equation}
\mathbf{H}_{BR}=A\left( \mathbf{I}_{x}\otimes \mathbf{S}_{x}+\mathbf{I}%
_{y}\otimes \mathbf{S}_{y}+\mathbf{I}_{z}\otimes \mathbf{S}_{z}\right)
+B\left[ a\mathbf{I}_{d}(2I+1)\otimes \mathbf{S}_{z}+b\mathbf{I}_{z}\otimes
\mathbf{I}_{d}(2)\right] ,  \label{eq:H_BR_matrix}
\end{equation}
where $\mathbf{I}_{d}(n)$ is the identity matrix of dimension $n$ and $%
\mathbf{I}_{u}$ and $\mathbf{S}_{u}$ are the well known nuclear and
electronic spin matrices. Obviously, for the electron we have the Pauli
matrices
\begin{equation}
\mathbf{S}_{x}=\frac{1}{2}\left(
\begin{array}{ll}
0 & 1 \\
1 & 0
\end{array}
\right) ,\;\mathbf{S}_{y}=\frac{1}{2}\left(
\begin{array}{ll}
0 & -i \\
i & 0
\end{array}
\right) ,\;\mathbf{S}_{z}=\frac{1}{2}\left(
\begin{array}{ll}
1 & 0 \\
0 & -1
\end{array}
\right) ,  \label{eq:spin_mat_1e}
\end{equation}
in units of $\hbar $. In what follows we consider two cases with different
nuclear spin.

In the simplest case $I=1/2$ the nuclear spin matrices are identical to (\ref
{eq:spin_mat_1e})
\begin{equation}
\mathbf{I}_{x}=\frac{1}{2}\left(
\begin{array}{ll}
0 & 1 \\
1 & 0
\end{array}
\right) ,\;\mathbf{I}_{y}=\frac{1}{2}\left(
\begin{array}{ll}
0 & -i \\
i & 0
\end{array}
\right) ,\;\mathbf{I}_{z}=\frac{1}{2}\left(
\begin{array}{ll}
1 & 0 \\
0 & -1
\end{array}
\right) ,  \label{eq:spin_mat_I_1/2}
\end{equation}
and the straightforward application of the Kronecker formula (\ref
{eq:H_BR_matrix}) leads to
\begin{equation}
\mathbf{H}_{BR}=\frac{1}{4}\left(
\begin{array}{llll}
A+2B\left( a+b\right) & 0 & 0 & 0 \\
0 & 2B(b-a)-A & 2A & 0 \\
0 & 2A & 2B(a-b)-A & 0 \\
0 & 0 & 0 & A-2B(a+b)
\end{array}
\right) .  \label{eq:H_BR_matrix_I_1/2}
\end{equation}
This matrix does not agree with those shown by Bhattacharya\cite{B07} and Oh
et al\cite{OHPK08} which also differ from each other. The three matrices
are, however, equivalent; for example $\mathbf{U}_{1}\mathbf{H}_{BR}\mathbf{U%
}_{1}^{T}=\mathbf{H}_{BR}^{B}$, where $\mathbf{H}_{BR}^{B}$ is
Bhattacharya's matrix and $\mathbf{U}_{1}$ is the orthogonal one
\begin{equation}
\mathbf{U}_{1}=\left(
\begin{array}{llll}
0 & 0 & 0 & 1 \\
0 & 0 & 1 & 0 \\
0 & 1 & 0 & 0 \\
1 & 0 & 0 & 0
\end{array}
\right) =\mathbf{U}_{1}^{T}=\mathbf{U}_{1}^{-1},  \label{eq:U_I_1/2_B}
\end{equation}
where $T$ stands for transpose. Analogously, the relation with the matrix of
Oh et al is $\mathbf{U}_{2}\mathbf{H}_{BR}\mathbf{U}_{2}^{T}=\mathbf{H}%
_{BR}^{OHPK}$, where
\begin{equation}
\mathbf{U}_{2}=\left(
\begin{array}{llll}
1 & 0 & 0 & 0 \\
0 & 0 & 1 & 0 \\
0 & 1 & 0 & 0 \\
0 & 0 & 0 & 1
\end{array}
\right) =\mathbf{U}_{2}^{T}=\mathbf{U}_{2}^{-1}.  \label{eq:U_I_1/2_OHPK}
\end{equation}
Clearly, $\mathbf{H}_{BR}$, $\mathbf{H}_{BR}^{B}$ and $\mathbf{H}%
_{BR}^{OHPK} $ are isospectral and, consequently, equivalent representations
of the same Hamiltonian $H_{BR}$. Note that the straightforward application
of the Kronecker product is equivalent to choosing a particular order in the
basis set (\ref{eq:basis_set}) which does not agree with the order chosen by
those other authors. The unitary transformations given by the matrices (\ref
{eq:U_I_1/2_B}) and (\ref{eq:U_I_1/2_OHPK}) are just two particular
permutations of the basis functions.

The matrix (\ref{eq:H_BR_matrix_I_1/2}) is the direct sum of two $1\times 1$
and one $2\times 2$ matrices. A matrix with this property is commonly called
block-diagonal and it is well known that its determinant is the product of
the determinants of each block. Therefore, the characteristic polynomial $%
P(E)=\det \left( \mathbf{H}_{BR}-E\mathbf{I}_{d}(4)\right) $has a
particularly simple form:

\begin{eqnarray}
P(E) &=&\frac{1}{256}\left[ 4E-A-2B\left( a+b\right) \right] \left[
4E-A+2B\left( a+b\right) \right] \times   \nonumber \\
&&\left[ 16E^{2}+8AE-3A^{2}-4B^{2}\left( a-b\right) ^{2}\right]
\label{eq:charpoly_I_1/2}
\end{eqnarray}
We realize that the problem of obtaining the four eigenvalues of $\mathbf{H}%
_{BR}$ reduces to solving a quadratic equation because two eigenvalues are
already known. The eigenvalues of this matrix were discussed in detail in
earlier papers\cite{B07,OHPK08}; here we are mainly interested in a simpler
construction of the matrix representation of the Hamiltonian operator.

\section{Second isotropic model}

\label{sec:isotropic2}

The next example is the case of a nucleus with spin $I=3/2$. The spin
matrices are given by

\begin{eqnarray}
\mathbf{I}_{x} &=&\frac{1}{2}\left(
\begin{array}{cccc}
0 & \sqrt{3} & 0 & 0 \\
\sqrt{3} & 0 & 2 & 0 \\
0 & 2 & 0 & \sqrt{3} \\
0 & 0 & \sqrt{3} & 0
\end{array}
\right) ,  \nonumber \\
\mathbf{I}_{y} &=&\frac{1}{2i}\left(
\begin{array}{cccc}
0 & \sqrt{3} & 0 & 0 \\
-\sqrt{3} & 0 & 2 & 0 \\
0 & -2 & 0 & \sqrt{3} \\
0 & 0 & -\sqrt{3} & 0
\end{array}
\right) ,  \nonumber \\
\mathbf{I}_{z} &=&\frac{1}{2}\left(
\begin{array}{cccc}
3 & 0 & 0 & 0 \\
0 & 1 & 0 & 0 \\
0 & 0 & -1 & 0 \\
0 & 0 & 0 & -3
\end{array}
\right) .  \label{eq:spin_mat_I_3/2}
\end{eqnarray}
Straightforward application of the Kronecker formula (\ref{eq:H_BR_matrix})
yields
\begin{landscape}
{\tiny
\begin{equation}
\mathbf{H}_{BR}=\frac{1}{4}\left(
\begin{array}{llllllll}
3A+2B(a+3b) & 0 & 0 & 0 & 0 & 0 & 0 & 0 \\
0 & 2B(3b-a)-3A & 2\sqrt{3}A & 0 & 0 & 0 & 0 & 0 \\
0 & 2\sqrt{3}A & A+2B(a+b) & 0 & 0 & 0 & 0 & 0 \\
0 & 0 & 0 & 2B(b-a)-A & 4A & 0 & 0 & 0 \\
0 & 0 & 0 & 4A & 2B(a-b)-A & 0 & 0 & 0 \\
0 & 0 & 0 & 0 & 0 & A-2B(a+b) & 2\sqrt{3}A & 0 \\
0 & 0 & 0 & 0 & 0 & 2\sqrt{3}A & 2B(a-3b)-3A & 0 \\
0 & 0 & 0 & 0 & 0 & 0 & 0 & 3A-2B(a+3b)
\end{array}
\right) .  \label{eq:H_BR_matrix_I_3/2}
\end{equation}
} \end{landscape}

This matrix differs from the one shown by Bhattacharya and Raman\cite{BR07}
in the order of some matrix elements. Both matrices are, however, equivalent
since they are related by the orthogonal matrix
\begin{equation}
\mathbf{U}=\left(
\begin{array}{llllllll}
0 & 0 & 0 & 0 & 0 & 0 & 0 & 1 \\
0 & 0 & 0 & 0 & 0 & 0 & 1 & 0 \\
0 & 0 & 0 & 0 & 0 & 1 & 0 & 0 \\
0 & 0 & 0 & 0 & 1 & 0 & 0 & 0 \\
0 & 0 & 0 & 1 & 0 & 0 & 0 & 0 \\
0 & 0 & 1 & 0 & 0 & 0 & 0 & 0 \\
0 & 1 & 0 & 0 & 0 & 0 & 0 & 0 \\
1 & 0 & 0 & 0 & 0 & 0 & 0 & 0
\end{array}
\right) .
\end{equation}
The matrix (\ref{eq:H_BR_matrix_I_3/2}) is the direct sum of two $1\times 1$
and three $2\times 2$ matrices and, consequently, its characteristic
polynomial can be written as the product of polynomials of smaller degree:

\begin{eqnarray}
P(E) &=&\frac{1}{65536}\left[ 4E-3A-2B\left( a+3b\right) \right] \left[
4e-3A+2B\left( a+3b\right) \right] \times  \nonumber \\
&&\left[ 16E^{2}+8AE-15A^{2}-4B^{2}\left( a-b\right) ^{2}\right] \times
\nonumber \\
&&\left[ 16E^{2}+8E\left( A+4Bb\right) -15A^{2}+4B\left( 2Aa+B\left(
a+b\right) \left( 3b-a\right) \right) \right] \times  \nonumber \\
&&\left[ 16E^{2}+8E\left( A-4Bb\right) -15A^{2}-4B\left( 2Aa+B\left(
a+b\right) \left( a-3b\right) \right) \right] .  \label{eq:charpoly_I_3/2}
\end{eqnarray}
Once again, the calculation of the eigenvalues of the matrix representation $%
\mathbf{H}_{BR}$ reduces to the calculation of the roots of quadratic
equations.

\section{Anisotropic model}

\label{sec:anisotropic}

In the examples above we have just considered isotropic cases, but the
application of the Kronecker-product approach to anisotropic models is
straightforward. For example, suppose that
\begin{equation}
H=\beta _{e}\mathbf{S}^{T}\cdot \mathbf{g}\cdot \mathbf{B}+\mathbf{S}%
^{T}\cdot \mathbf{A}\cdot \mathbf{I}-\beta _{n}\mathbf{I}^{T}.\mathbf{g}_{n}.%
\mathbf{B},  \label{eq:H_aniso_1}
\end{equation}
where $\mathbf{S}$ and $\mathbf{I}$ are column matrices with components $%
S_{x}$, $S_{y}$, $S_{z}$ and $I_{x}$, $I_{y}$ and $I_{z}$, respectively, and
$\mathbf{g}$, $\mathbf{A}$, and $\mathbf{g}_{n}$ are $3\times 3$ matrices%
\cite{DW91}. In order to apply the Kronecker product we rewrite it as
\begin{equation}
H=a_{1}I_{x}+a_{2}I_{y}+a_{3}I_{z}+b_{1}S_{x}+b_{2}S_{y}+b_{3}S_{z}+c_{11}I_{x}S_{x} +c_{12}I_{x}S_{y}+c_{13}I_{x}S_{z}+c_{23}I_{y}S_{z},
\label{eq:H_aniso_2}
\end{equation}
where we have omitted the identity operators in the linear terms; for
example $I_{u}\otimes \hat{1}_{S}$ or $\hat{1}_{I}\otimes S_{u}$ (see our
earlier paper for more details\cite{F16}).

The matrix representation of the Hamiltonian can then be easily obtained
from
\begin{eqnarray}
\mathbf{H}_{A} &=&a_{1}\mathbf{I}_{x}\otimes \mathbf{I}_{d}(2)+a_{2}\mathbf{I%
}_{y}\otimes \mathbf{I}_{d}(2)+a_{3}\mathbf{I}_{z}\otimes \mathbf{I}_{d}(2)
\nonumber \\
&&+b_{1}\mathbf{I}_{d}(2)\otimes \mathbf{S}_{x}+b_{2}\mathbf{I}%
_{d}(2)\otimes \mathbf{S}_{y}+b_{3}\mathbf{I}_{d}(2)\otimes \mathbf{S}_{z}
\nonumber \\
&&+c_{11}\mathbf{I}_{x}\otimes \mathbf{S}_{x}+c_{12}\mathbf{I}_{x}\otimes
\mathbf{S}_{y}+c_{13}\mathbf{I}_{x}\otimes \mathbf{S}_{z}+c_{23}\mathbf{I}%
_{y}\otimes \mathbf{S}_{z}  \label{eq:H_mat_aniso_1}
\end{eqnarray}
The result is the Hermitian matrix {\tiny
\begin{equation}
\mathbf{H}_{A}=\frac{1}{4}\left(
\begin{array}{llll}
2\left( a_{3}+b_{3}\right) & 2b_{1}-2ib_{2} & 2a_{1}+c_{13}-i\left(
2a_{2}+c_{23}\right) & c_{11}-ic_{12} \\
2b_{1}+2ib_{2} & 2\left( a_{3}-b_{3}\right) & c_{11}+ic_{12} &
2a_{1}-c_{13}+i\left( c_{23}-2a_{2}\right) \\
2a_{1}+c_{13}+i\left( 2a_{2}+c_{23}\right) & c_{11}-ic_{12} & 2\left(
b_{3}-a_{3}\right) & 2b_{1}-2ib_{2} \\
c_{11}+ic_{12} & 2a_{1}-c_{13}+i\left( 2a_{2}-c_{23}\right) & 2b_{1}+2ib_{2}
& -2\left( a_{3}+b_{3}\right)
\end{array}
\right) ,  \label{eq:H_mat_aniso_2}
\end{equation}
}that contains the matrix (\ref{eq:H_BR_matrix_I_1/2}) as a particular case.
In this case the matrix $\mathbf{H}_{A}-E\mathbf{I}_{d}(4)$ is not
block-diagonal and the characteristic polynomial will not exhibit a simple
form; consequently, we will have to solve the characteristic polynomial of
fourth order. However, the point is that the construction of the matrix
representation of the Hamiltonian (\ref{eq:H_aniso_2}) is greatly
facilitated by the straightforward application of the Kronecker product.

\section{Conclusions}

\label{sec:conclusions}

The aim of this sequel of our earlier paper\cite{F16} is to show that one
can easily obtain the matrix representation of a wide variety of spin
Hamiltonians without resorting to a basis set and ladder operators for the
calculation of the matrix elements. The straightforward application of the
Kronecker product yields the desired result. To this end it is only
necessary to have the matrix representations of the spin matrices of all the
particles in the Hamiltonian operator. This approach is particularly
appealing if one has a suitable program for the calculation of the Kronecker
product within a computer algebra system. In the present case we resorted to
the computer algebra system Derive (https://education.ti.com/en/us/home) and
one of the Kronecker-product programs contributed to the Derive User Group
(http://www.austromath.at/dug/). The examples discussed in the preceding
sections clearly illustrate the remarkable simplicity of the
Kronecker-product method. Note that we used basically the same formula for
the isotropic cases with $I=1/2$ and $I=3/2$ and we easily adapted it to the
anisotropic case with $I=1/2$.

The spin matrices necessary for the application of the Kronecker product are
easily obtained by means of the commutation properties of the spin
operators. The analytic expressions for the matrix elements are well known
and therefore available from several sources (see, for example,
http://easyspin.org/documentation/spinoperators.html).

\end{document}